\documentclass[a4paper,11pt]{article}
\pdfoutput=1 

\usepackage{jcappub}

\usepackage[T1]{fontenc}

\usepackage{graphicx}
\title{\boldmath Erratum: Nonlinear spherical perturbations in quintessence models of dark energy} 

\author[a]{Manvendra Pratap Rajvanshi,}
\author[a,1]{J. S. Bagla,\note{Corresponding author}}

\affiliation[a]{IISER Mohali, Sector 81, Sahibzada Ajit Singh Nagar,
  Punjab 140306, India} 

\emailAdd{manvendra@iisermohali.ac.in}
\emailAdd{jasjeet@iisermohali.ac.in}

\begin{document}

\maketitle

\section{Introduction}

We reported results of our study on non-linear spherical perturbations
in quintessence models of dark energy.
In the process of some follow up studies we discovered that a scaling
factor in the code used for numerical calculations that should have
been set to unity was set to a large value ($10^3$).
Thus the scale of perturbations was much larger than intended, and for
the larger scales the amplitude of dark matter perturbations was much
higher than realistic.
We provide corrected results here in this erratum.
We find that there is no change in the perturbations for dark matter.
The amplitude of perturbations in dark energy is much smaller than
presented in the paper \cite{2018JCAP.06.018}.
Same holds true for spatial variation in the equation of state
parameter.

Our key conclusions do not change.
\begin{itemize}
\item
  Dark energy perturbations do not lead to any variation in dark
  matter perturbations, e.g., radius at turn around and the virial
  radius, as well as density contrast at these two epochs remains
  unchanged.
\item
  Dark energy perturbations grow at the linear rate, or faster.
\item
  Dark energy perturbations remain small at all times and scales in
  response to non-linear dark matter perturbations.
\item
  Equation of state parameter $w$ becomes a function of scale and the
  variation is related to the density contrast in dark matter.
\item
  Gradients in dark energy and metric coefficients are strongly
  suppressed.
\end{itemize}

We find that two of the approaches discussed and discarded in the
original paper do not work properly with the correct scaling.
This is described in some detail.

For completeness, we give here the corrected figures, where the
incorrect scaling factor in the code gave us incorrect variation. 
\textbf{Note}: We have indexed the figures in this erratum to have
same figure numbers as in original article \cite{2018JCAP.06.018}. Two
extra figures are added to compliment figure 8 and 13, these are
labeled as figure 8(extra) and figure 13(extra), respectively. 

\section{Dark energy in over dense profile}

The initial profile for dark matter perturbation is taken to be(see
section 2.0.1 of article \cite{2018JCAP.06.018} for details): 
\begin{equation}
\delta_i(r) = \left\{ \begin{array}{lll} \alpha_0
  \left[1-\left(\frac{r}{\sigma_0}\right)^2 
  \right]^2 -\alpha_1  \left[1-\left(\frac{r}{\sigma_1}\right)^2
  \right]^2 & {\,\,\,\,} & \left(r\leq\sigma_0\right)  \\
 -\alpha_1  \left[1-\left(\frac{r}{\sigma_1}\right)^2 \right]^2 & {\,\,\,\,} &
 \left(\sigma_0 < r\leq\sigma_1\right) \\
 0   & {\,\,\,\,} & \left(r>\sigma_1\right) \\
\end{array}
\right.
\end{equation}
For the over dense case the scales used are small
($\sigma_0=3,\sigma_1=18$, $(3,18)$) and hence dark energy 
perturbations are very weak. 
The claim that spatio-temporal fluctuations develop in dark
energy continue to hold, though the amplitude of these fluctuations is
much smaller than was reported earlier.
This is very clear for the under dense case which has a larger scale:
we had shown that dark energy perturbations at larger scales develop a
larger amplitude for the same amplitude of dark matter perturbations. 

We find that after the corrections, dark energy perturbations are very
small at small scales, e.g. for $(3,18)$ mentioned above.
Thus we supplement the corrected figures with results for a
super cluster size halo with two length parameters as
$(30,75)$~Mpc(figures in \ref{fig:7370}).
This case shows a clear evolution of spatio-temporal perturbations.
Density contrast in dark energy at different redshifts is shown in
fig~\ref{fig:9}. 

\setcounter{figure}{7}
\begin{figure}[hbt]
    \centering
    \includegraphics[width=0.45\textwidth]{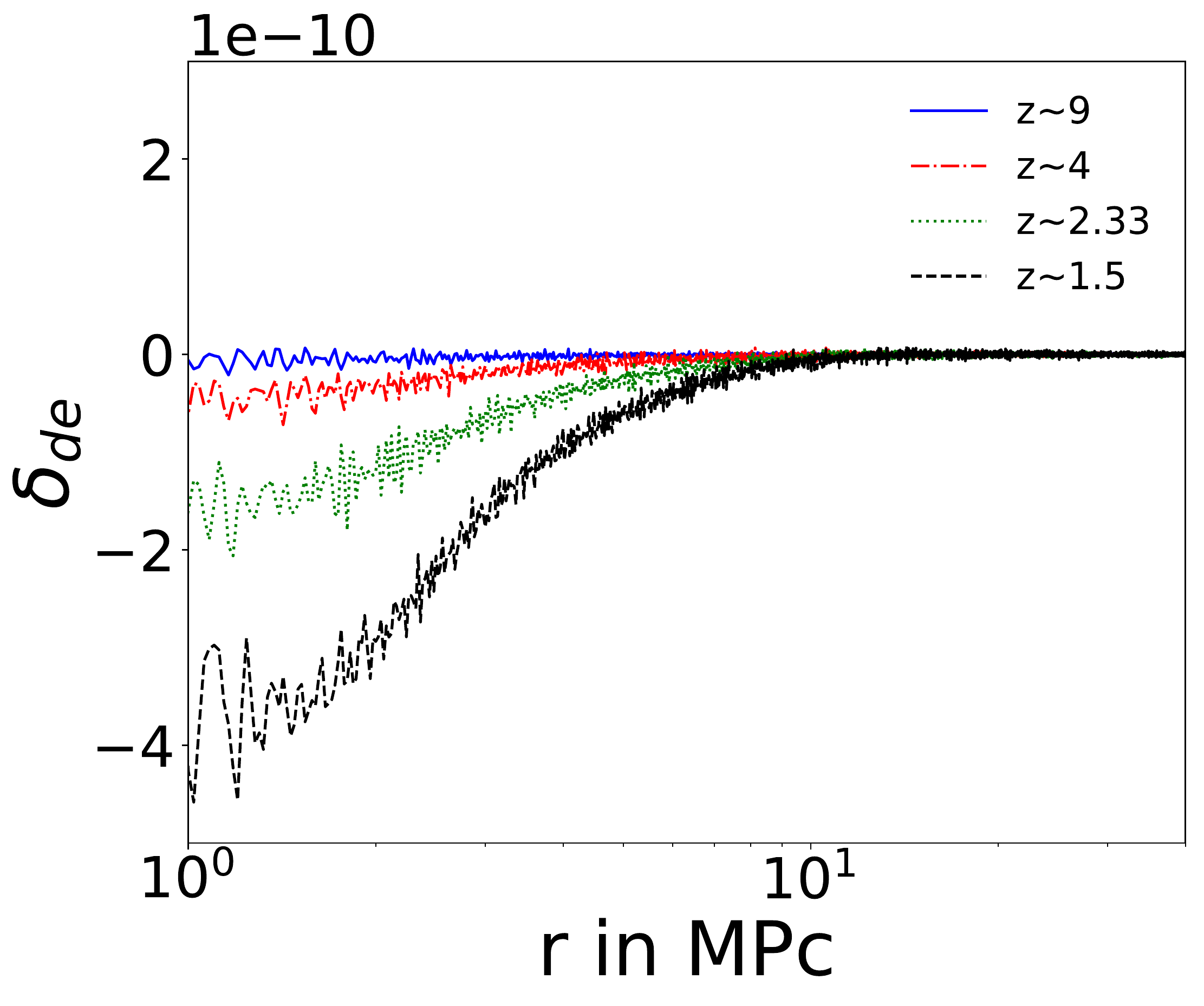}
    \includegraphics[width=0.45\textwidth]{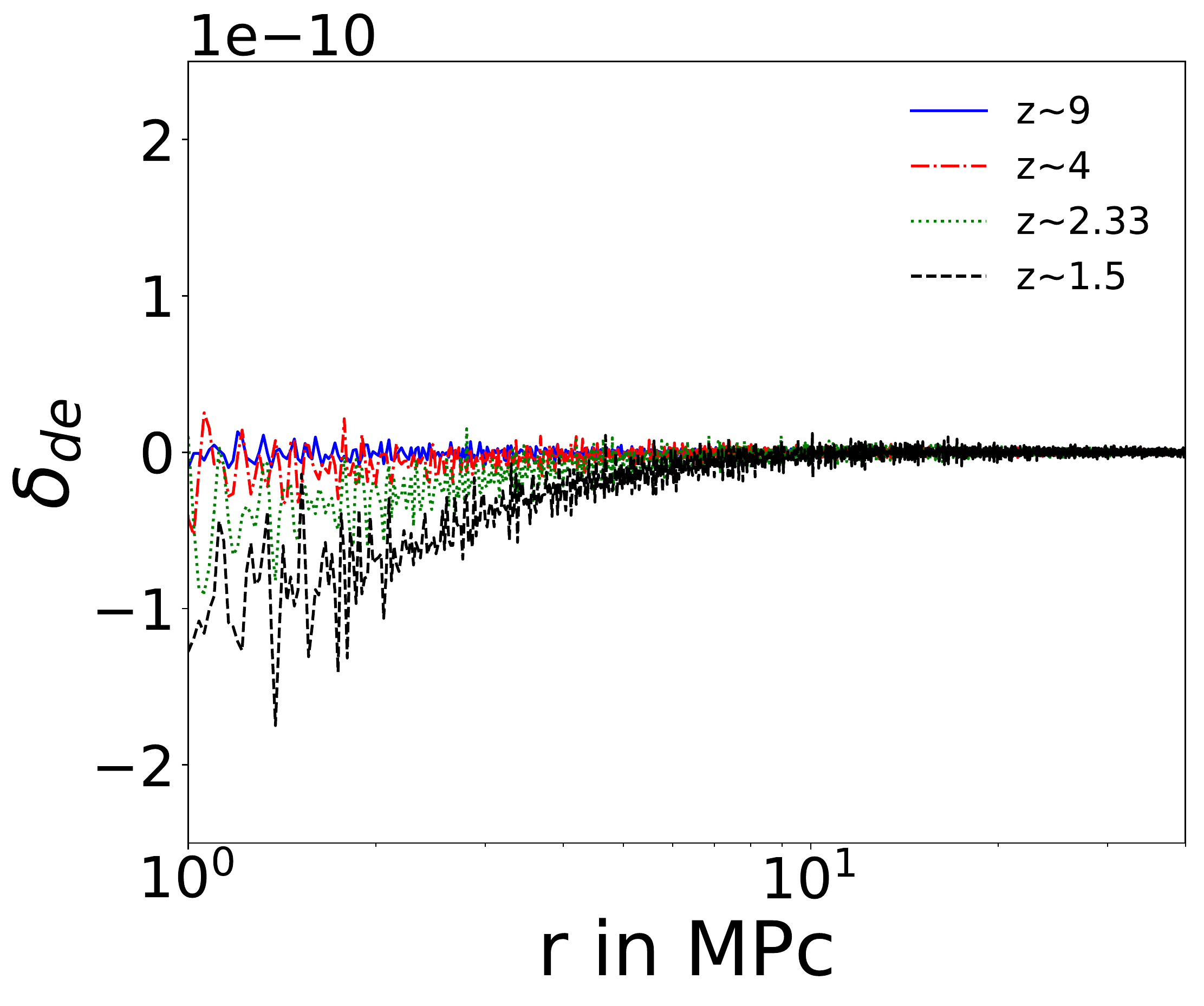}
    \caption{Density contrast for dark energy as a function of scale
      at different epochs.  We see that the amplitude of perturbations
      in dark energy remains small at all scales and at all times.
      The left panel is for $V\propto \psi^2 $ while the right panel
      is for $V\propto \exp(-\psi) $.}
    \label{fig:9}
\end{figure}

\begin{figure}[hbt]
\renewcommand{\thefigure}{8(extra)}
    \centering
    \includegraphics[width=0.45\textwidth]{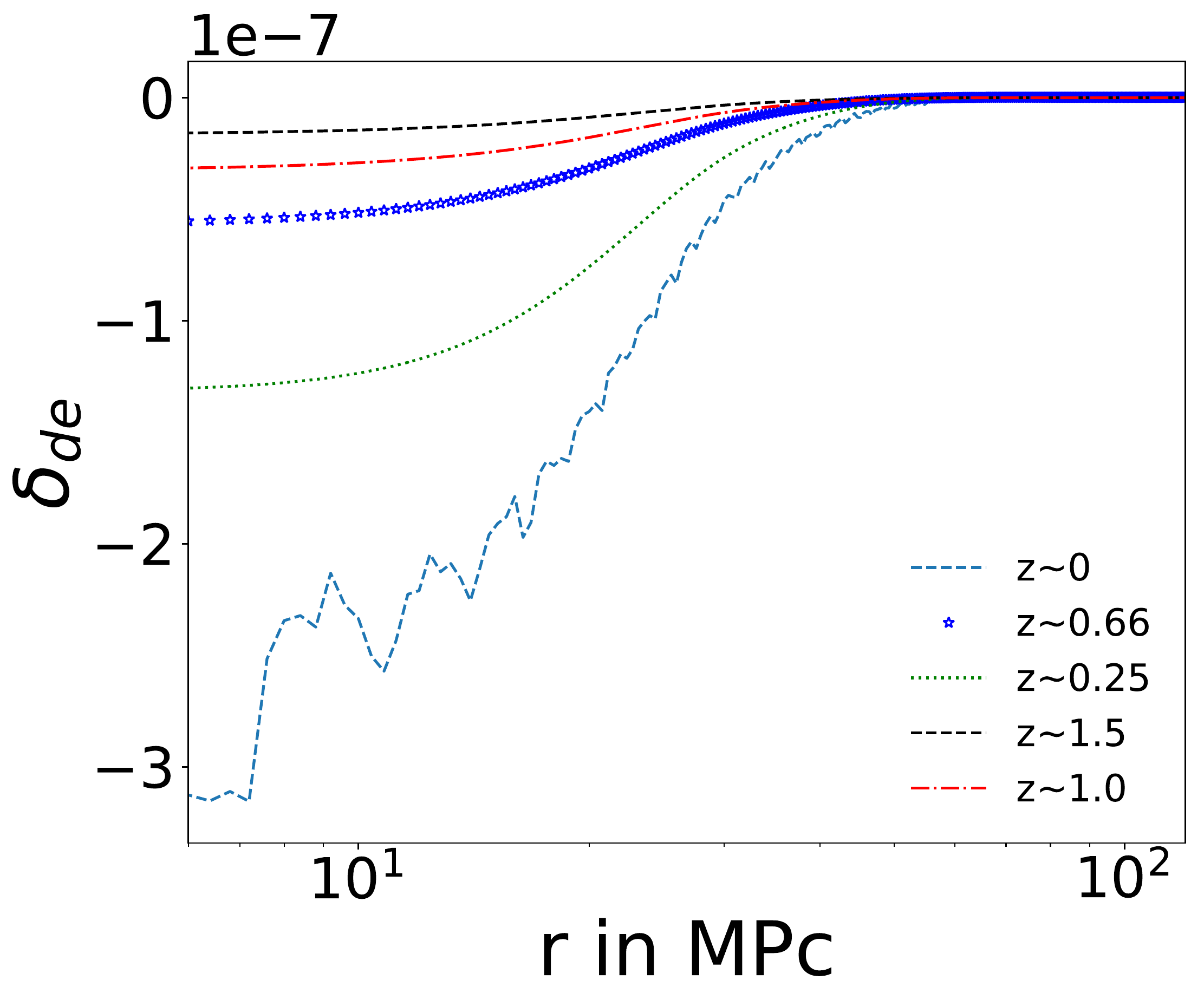}
    \includegraphics[width=0.45\textwidth]{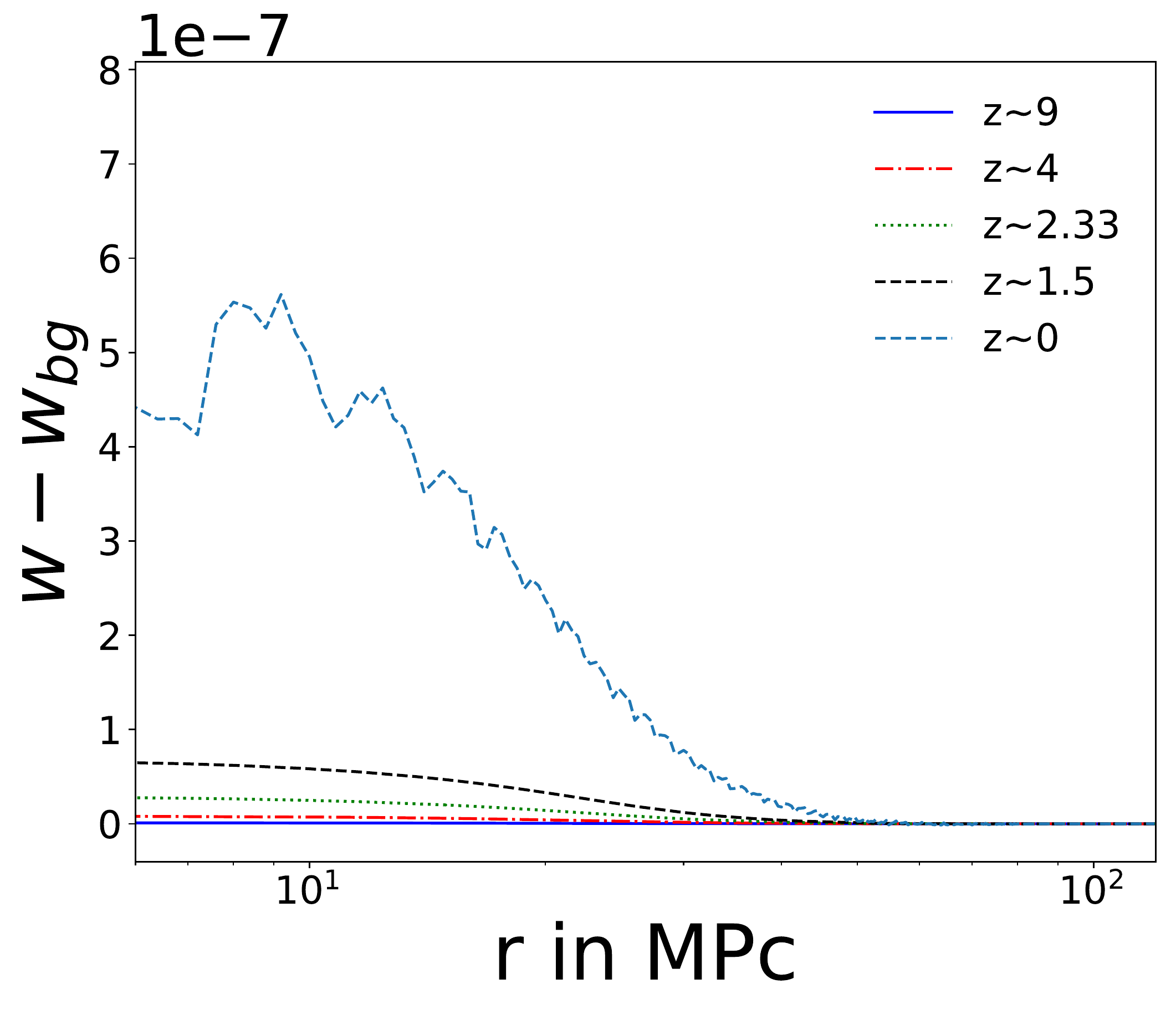}
    \caption{Density contrast for dark energy for a super cluster
      sized perturbation collapsing around $z \simeq 0.02$.}
    \label{fig:7370}
\end{figure}

Evolution of the equation of state for dark energy $w$ is shown in
\ref{fig:8}.
We have plotted the difference between the value of $w$ and the
expected value in the {\sl background} model, taken here to be the
value at large scales in the simulation.
We can see that the fluctuations are non-zero but very small.

\setcounter{figure}{8}
\begin{figure}[hbt]
    \centering
    \includegraphics[width=0.45\textwidth]{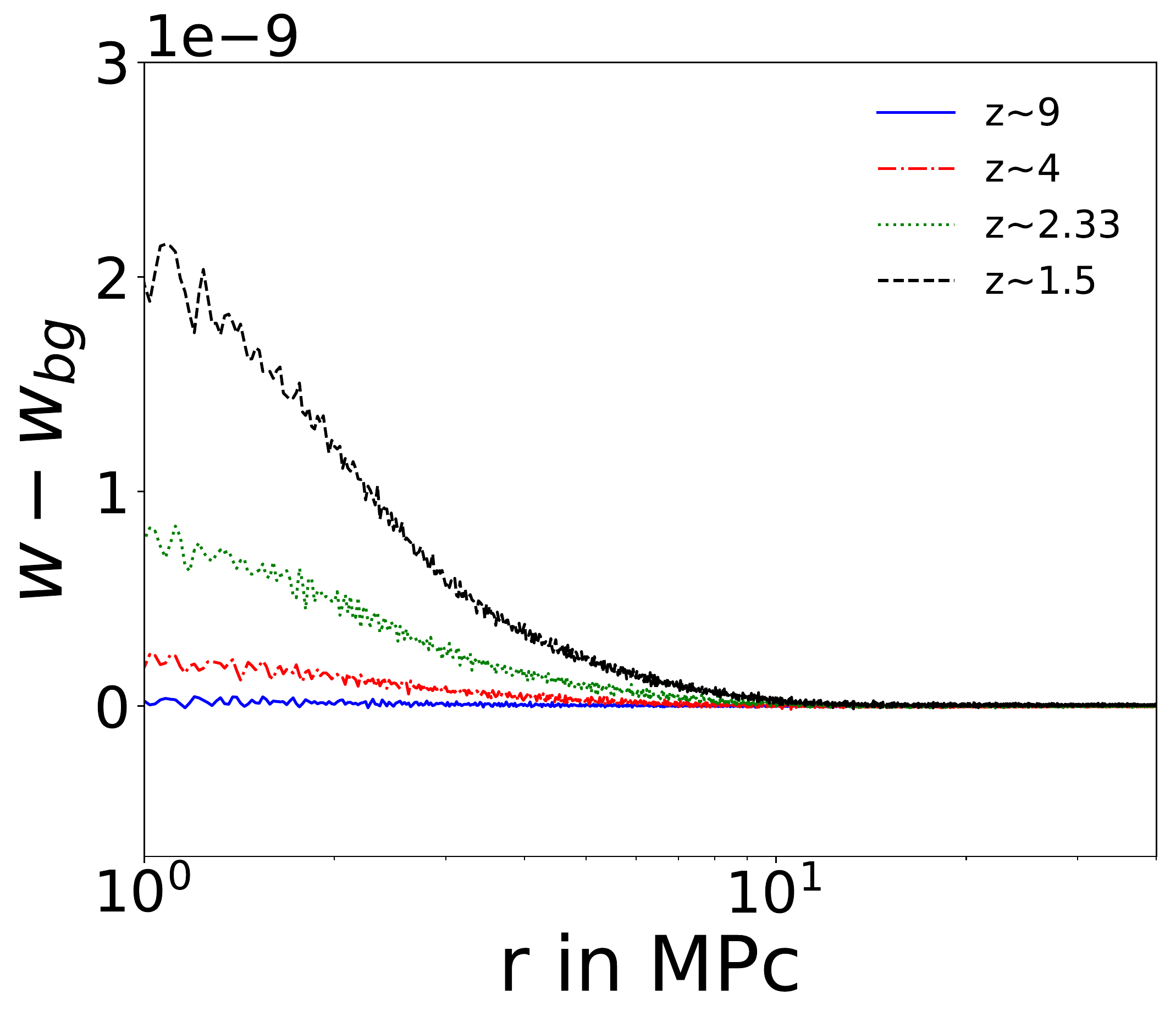}
    \includegraphics[width=0.5\textwidth]{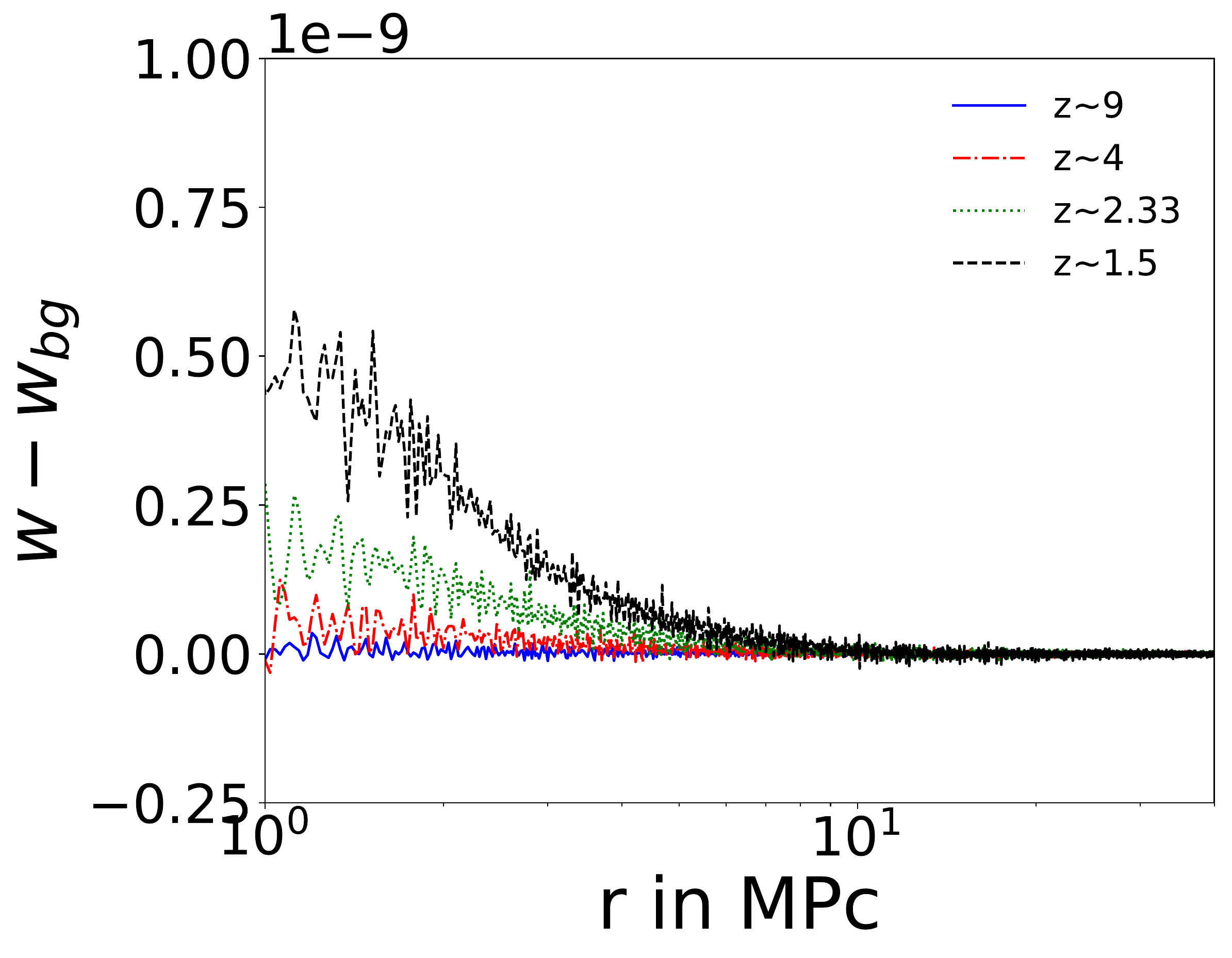}
    \caption{ The equation of state parameter for dark
      energy as a function of scale at different epochs.}
    \label{fig:8}
\end{figure}

\section{Dark Energy in under dense regions}

\begin{figure}[hbt]
    \centering
    \includegraphics[width=0.45\textwidth]{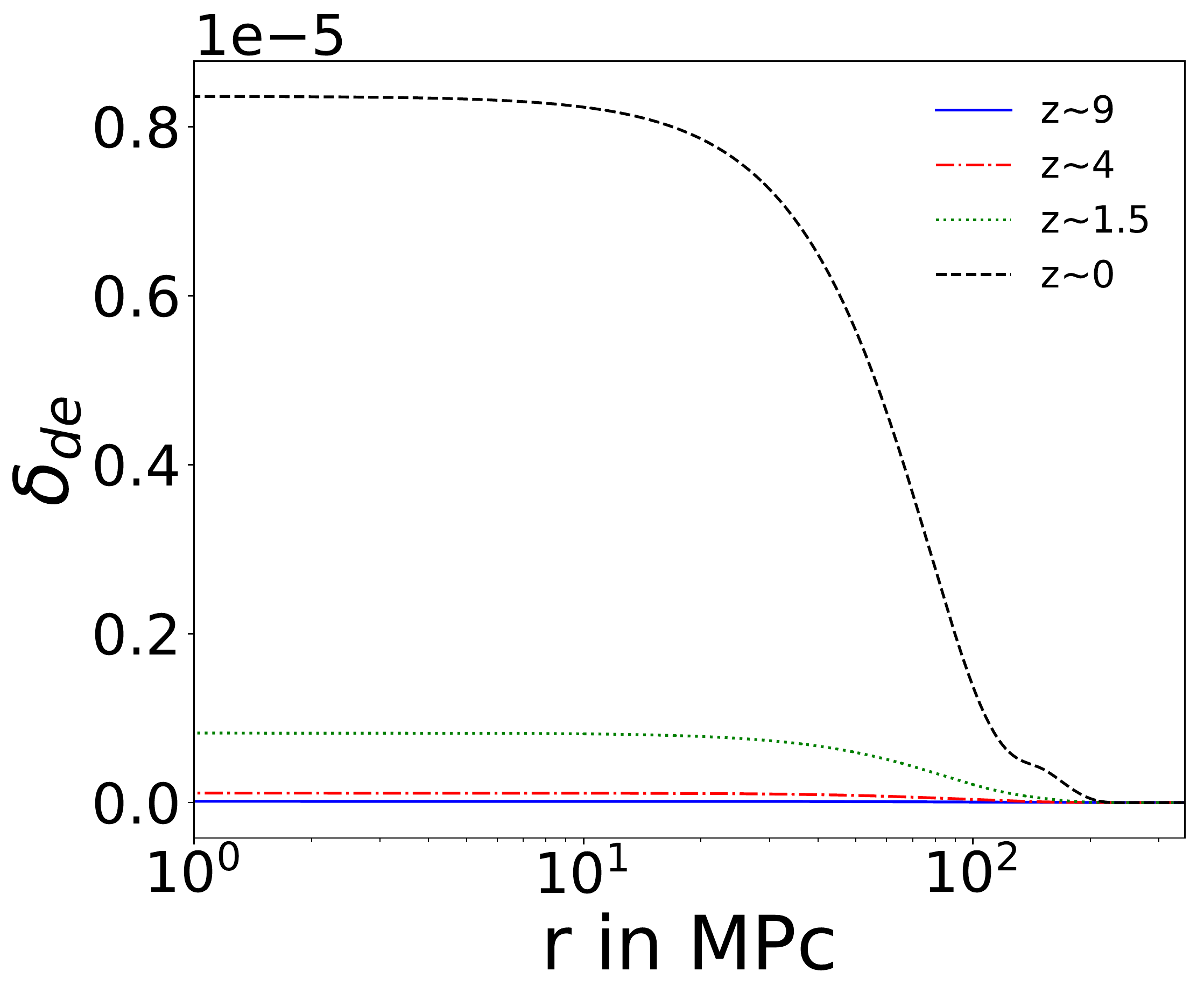}
    \includegraphics[width=0.45\textwidth]{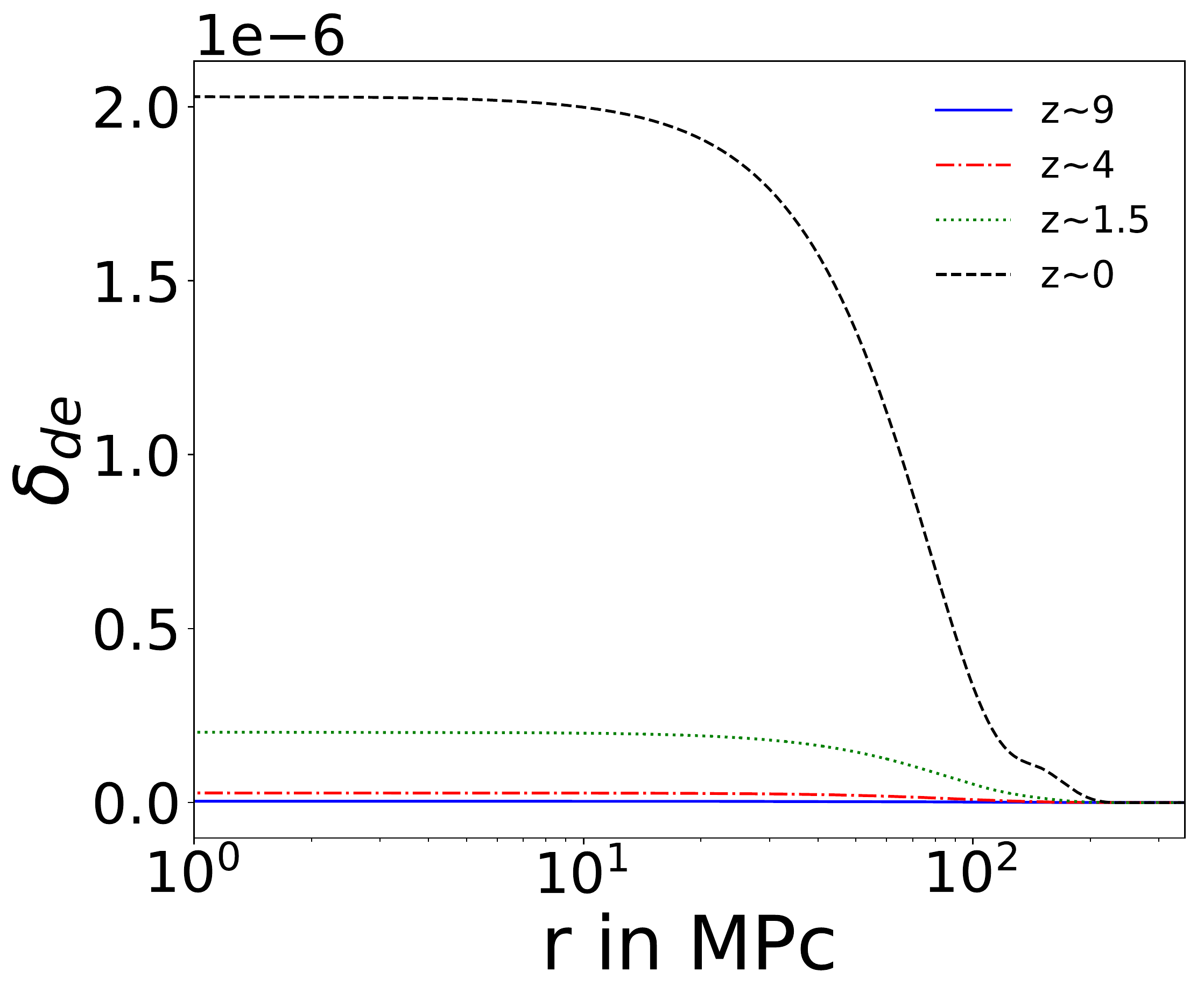}
    \caption{Density contrast for dark energy as a function
  of scale $r$ for a matter under-density from simulation
  UD1.  This is
  plotted at multiple epochs.  We find that dark energy perturbations
  grow but the amplitude remains small in absolute terms.   The left
  panel is for $V\propto \psi^2 $ while the right panel is for
  $V\propto \exp(-\psi) $.}
    \label{fig:5a}
\end{figure}

\begin{figure}[hbt]
    \centering
    \includegraphics[width=0.45\textwidth]{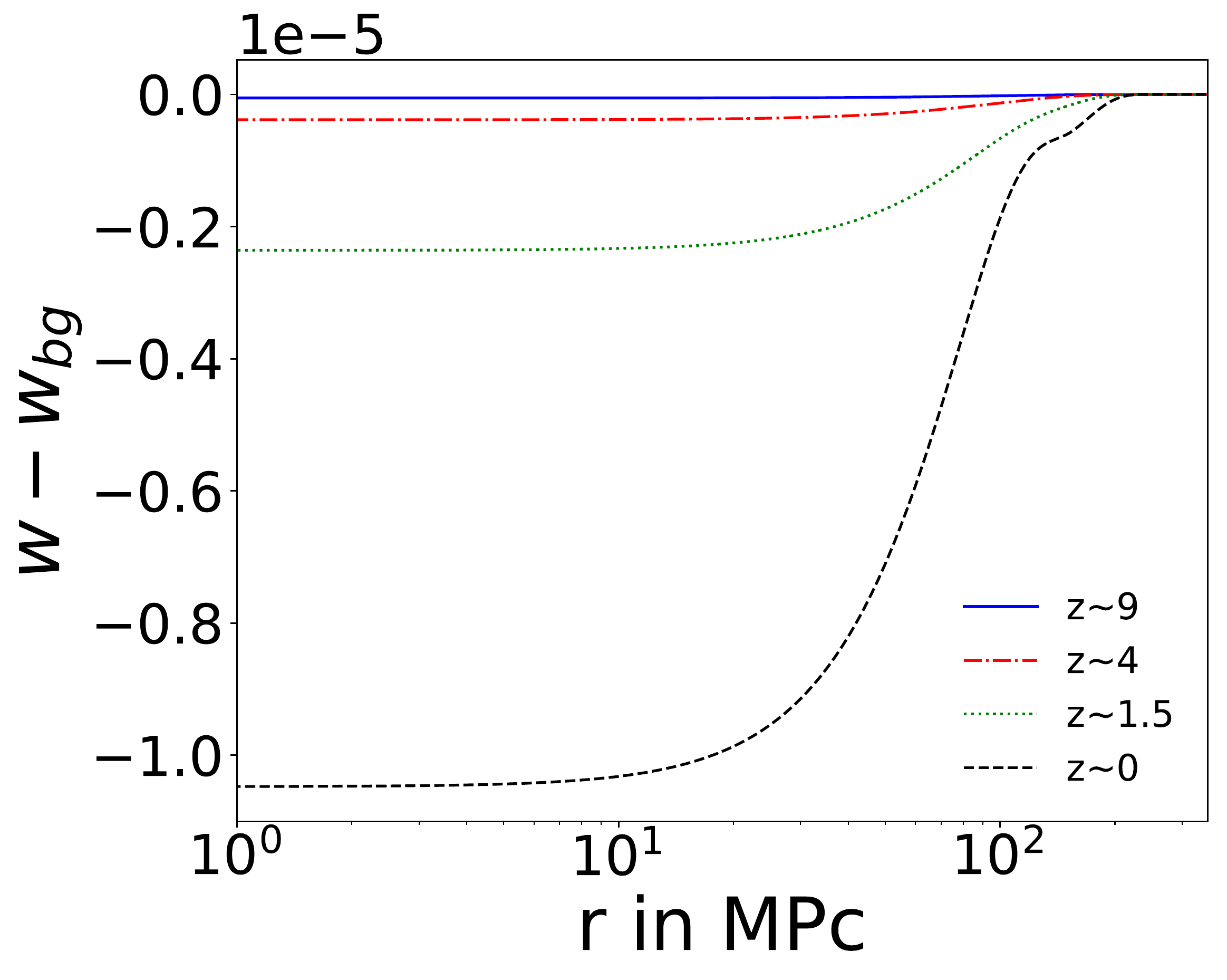}
    \includegraphics[width=0.45\textwidth]{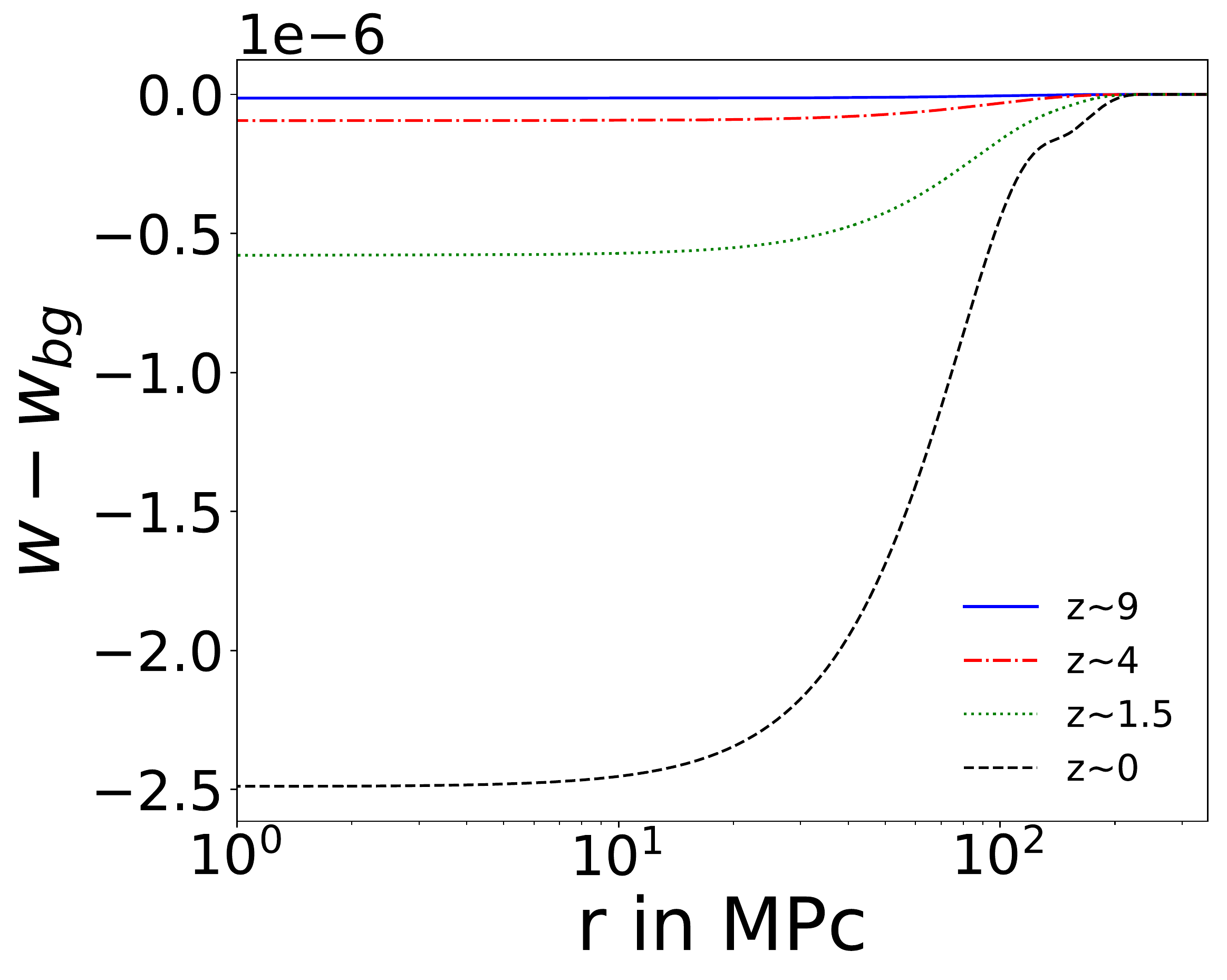}
    \caption{ Equation of state parameter $w$ as a function
  of scale $r$ for a void, i.e., a matter under-density for simulation
  UD1.  This is
  plotted at multiple epochs.  We find that $w$ inside the void is
  smaller than at large scales.  The left
  panel is for $V\propto \psi^2 $ while the right panel is for
  $V\propto \exp(-\psi) $.}
    \label{fig:5b}
\end{figure}

We present the variation of density contrast for dark energy and the
equation of state parameter $w$ in figures \ref{fig:5a} and
\ref{fig:5b}.
The amplitude of fluctuations is much smaller than presented in the
original work. 
The variation of $w$ with time is much more significant than the
spatial variation.

\section{Comparison with linear theory}

\begin{figure}[hbt]
    \centering
    \includegraphics[width=0.45\textwidth]{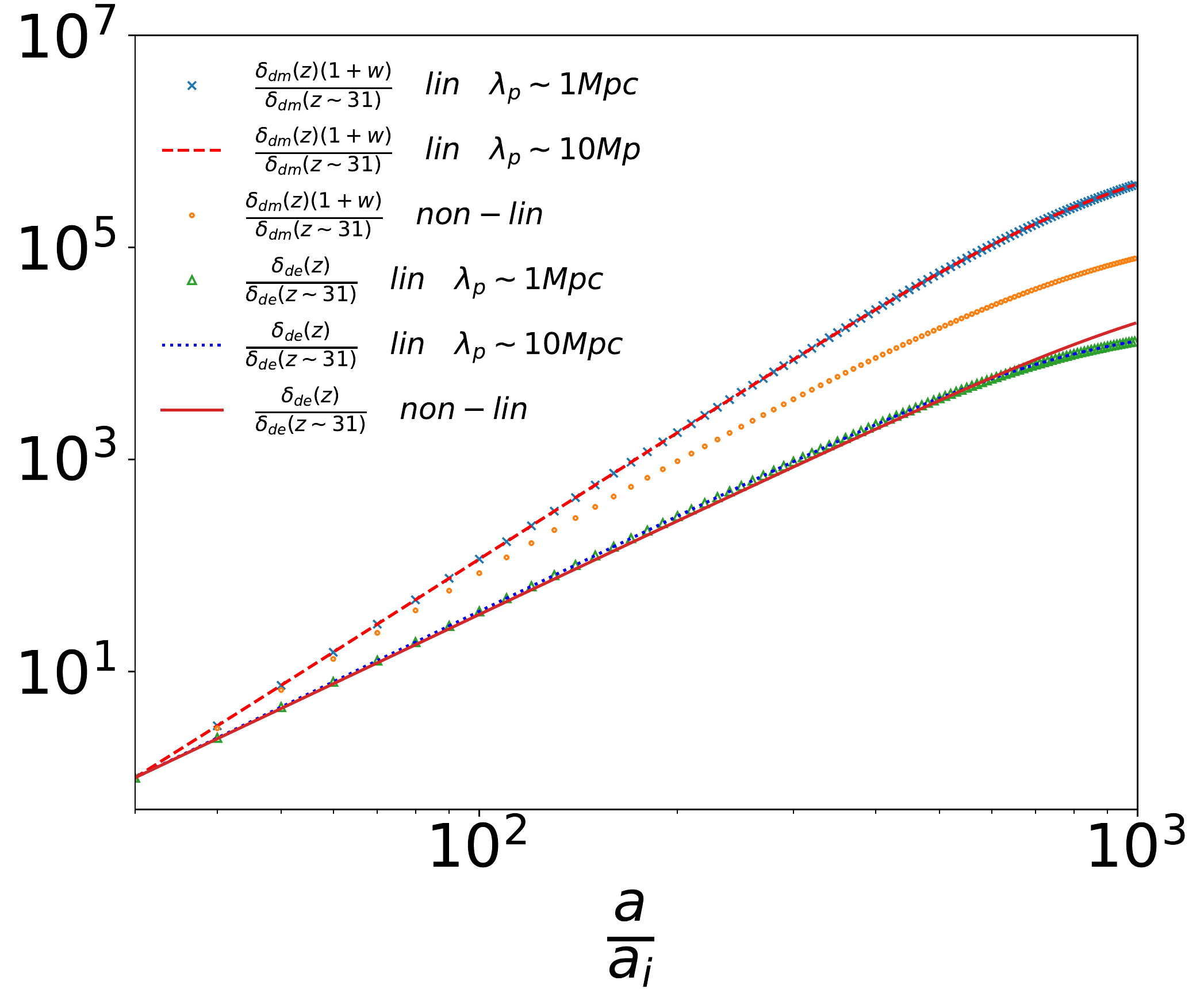}
    \includegraphics[width=0.45\textwidth]{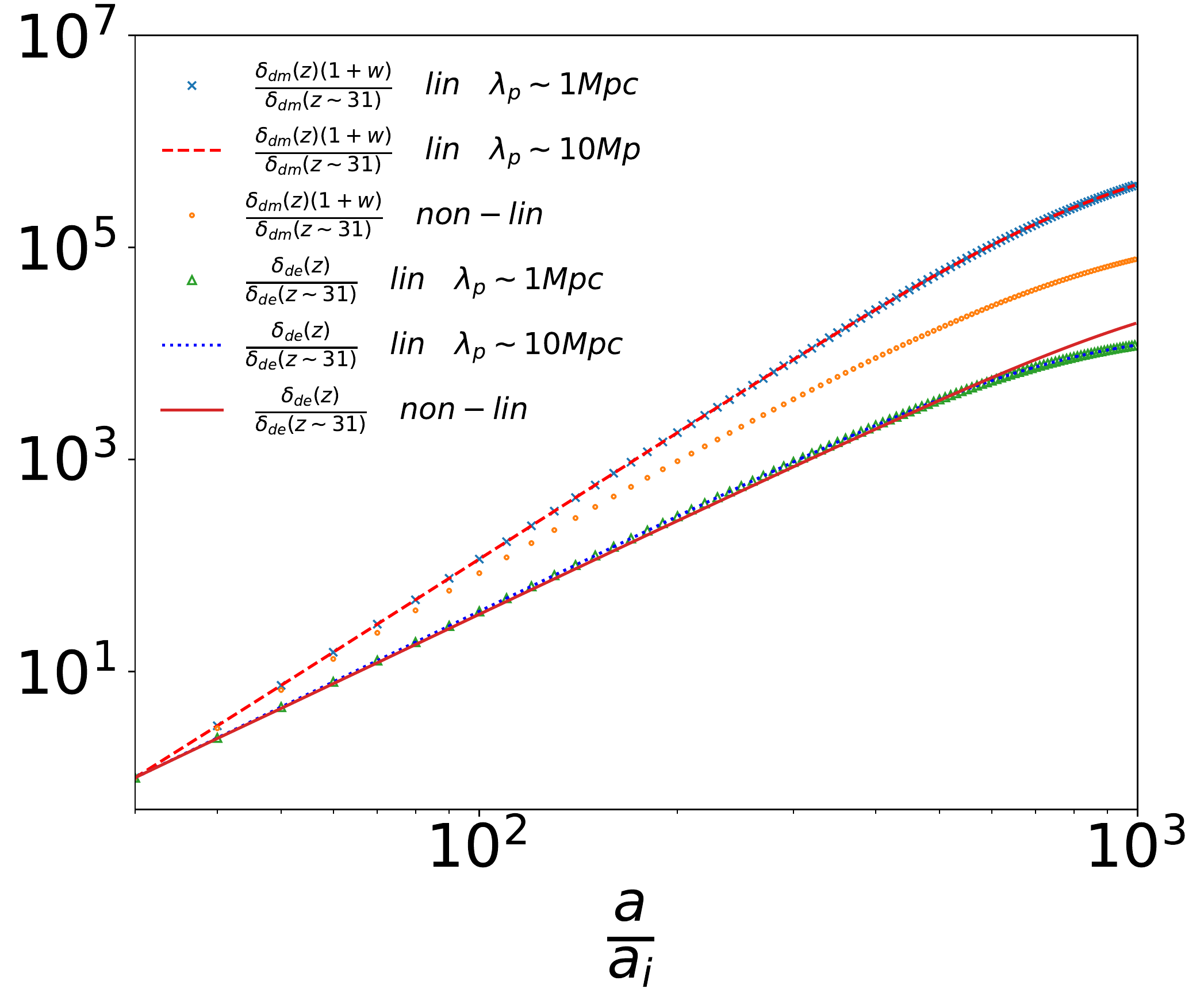}
    \caption{A comparison of the evolution of dark energy
  perturbations.
  Left panel is for $V\propto \psi^2 $ while right one are for
  $V\propto \exp(-\psi) $.
  At very large scales the linear theory prediction for the
  magnitude of dark energy perturbations scales as $(1+w)\delta_{dm}$.
  We have plotted this combination for linearly evolved $\delta_{dm}$
  for two scales: $1$~Mpc (cross) and $10$~Mpc (dashed line).  Linear
  evolution of dark energy density contrast for the two scales is also
  shown here as triangles ($1$~Mpc) and dotted line ($10$~Mpc).  We
  find that the linear evolution for dark energy perturbations is
  slower at small scales as compared to the expected variation at
  large scales.
  All points pertaining to linear evolution are normalised to unity at
  the left corner.}
    \label{fig:11}
\end{figure}

The conclusion presented in article that nonlinear perturbations at
later times grow faster than linear ones continues to hold for
corrected scales.
The correspondence between growth rate of $\delta_{dm}(1+w)$ and
growth rate of $\delta_{de}$ holds through and is similar as simulated
earlier.
The corrected variation is presented in fig~\ref{fig:11}.

\section{Role of spatial gradient in field dynamics}

In this context the results as reported in fig~13 of
\cite{2018JCAP.06.018} holds and here we plot the graph with corrected
scales in fig~(\ref{fig:14ofa}).

\setcounter{figure}{12}
\begin{figure}
  \renewcommand{\thefigure}{13}
  \centering
    \includegraphics[width=0.45\textwidth]{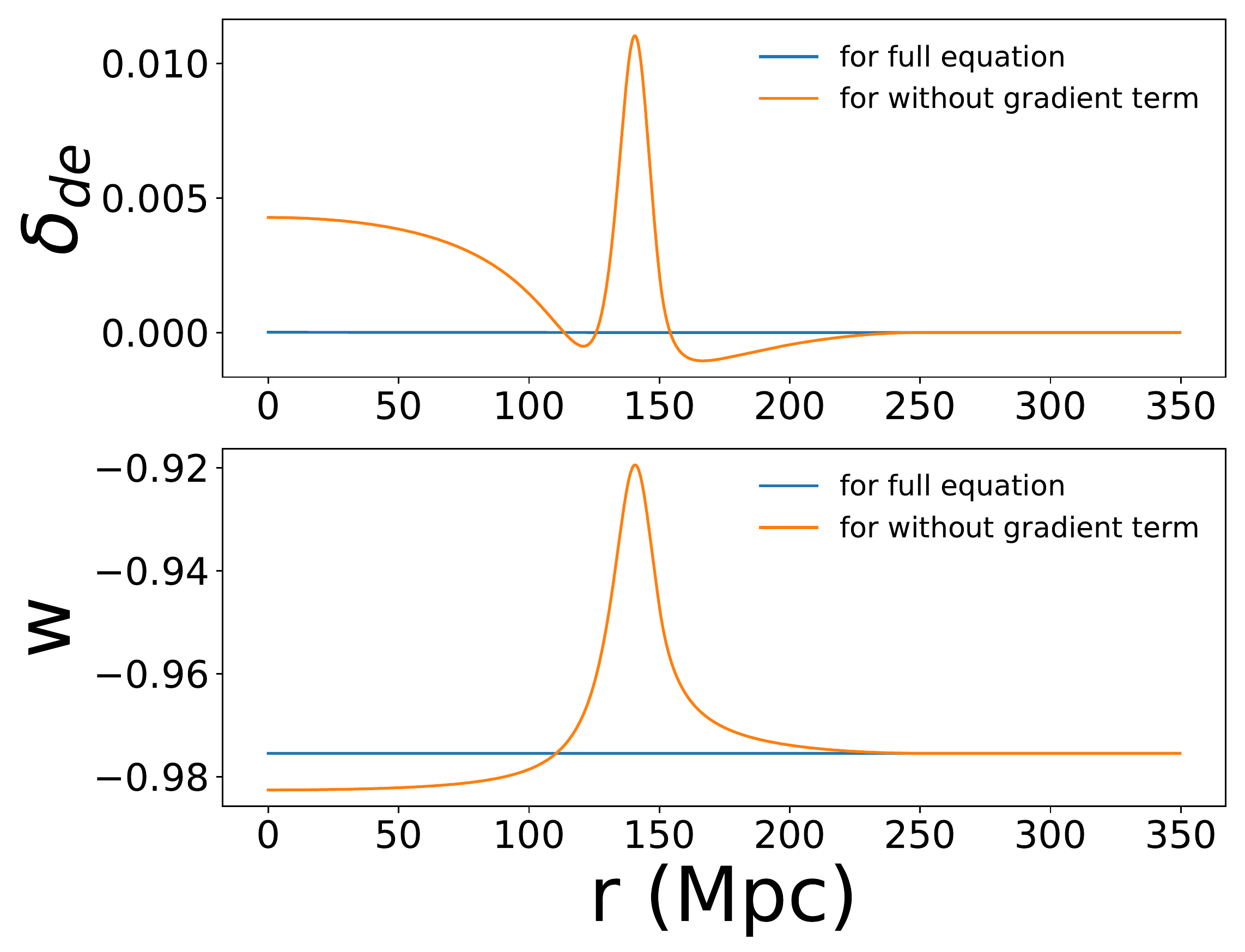}
    \includegraphics[width=0.45\textwidth]{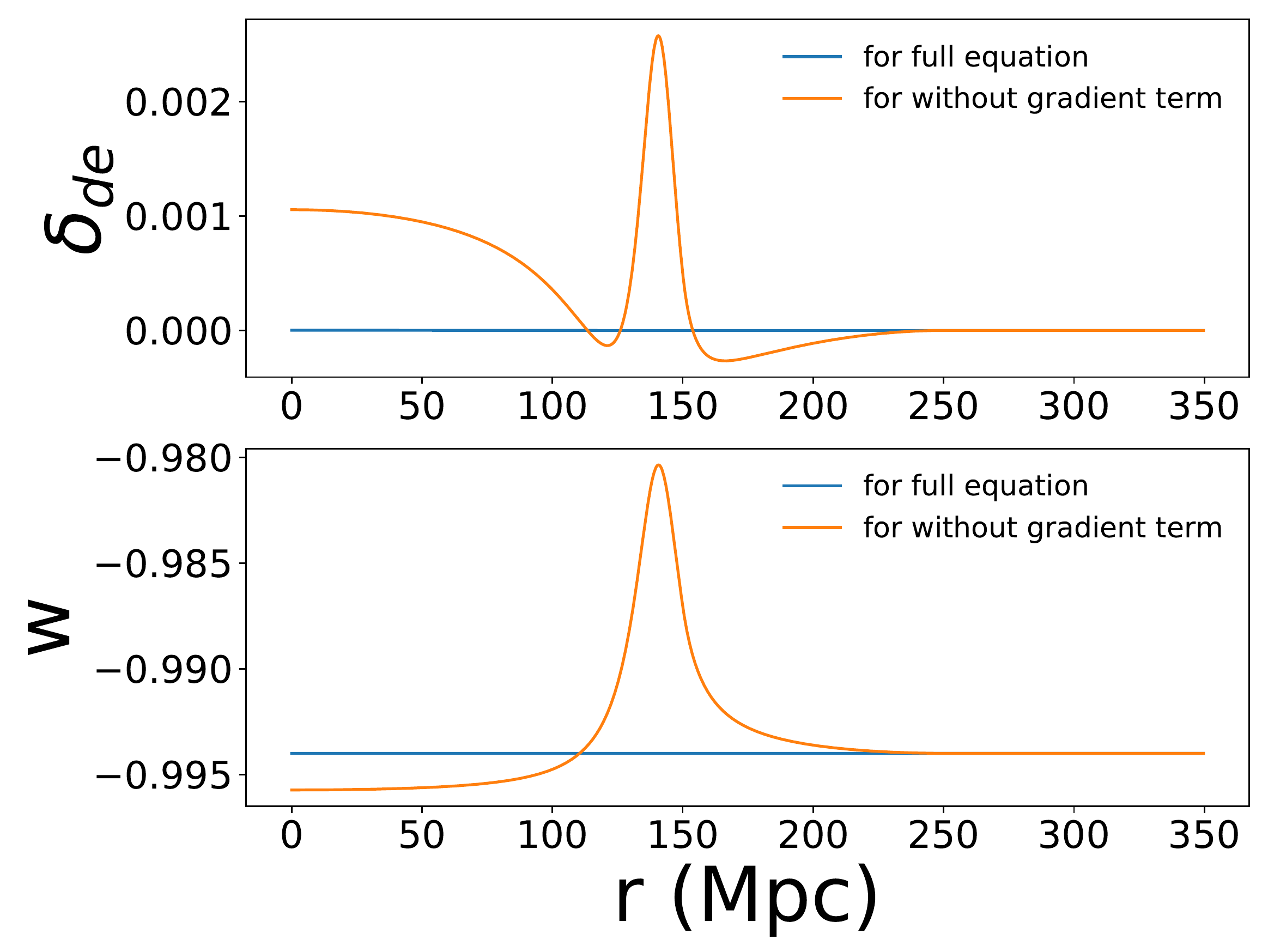}
    \caption{In this figure we explore the leading cause of variation
      of equation of state parameter $w$ for simulation UD1.
      We show the variation computed by retaining only the local
      Hubble expansion terms in the equation of motion and compare it
      with the full simulation.
      In the former case, we ignore the gradient term.
      We find that the variation of $w$ is fairly strong and has some
      localised features when the gradient terms are ignored.
      The localised features are not present in the full simulation
      indicating that the gradients of the scalar field are suppressed
      in the evolution, and the local Hubble expansion is not the only
      determining factor.
      This result remains valid after correction in the scale, though
      with a reduced amplitude of fluctuations.}
    \label{fig:14ofa}
\end{figure}

\section{Effect of scales}

\begin{figure}
\renewcommand{\thefigure}{14}
    \centering
     \includegraphics[width=0.45\textwidth]{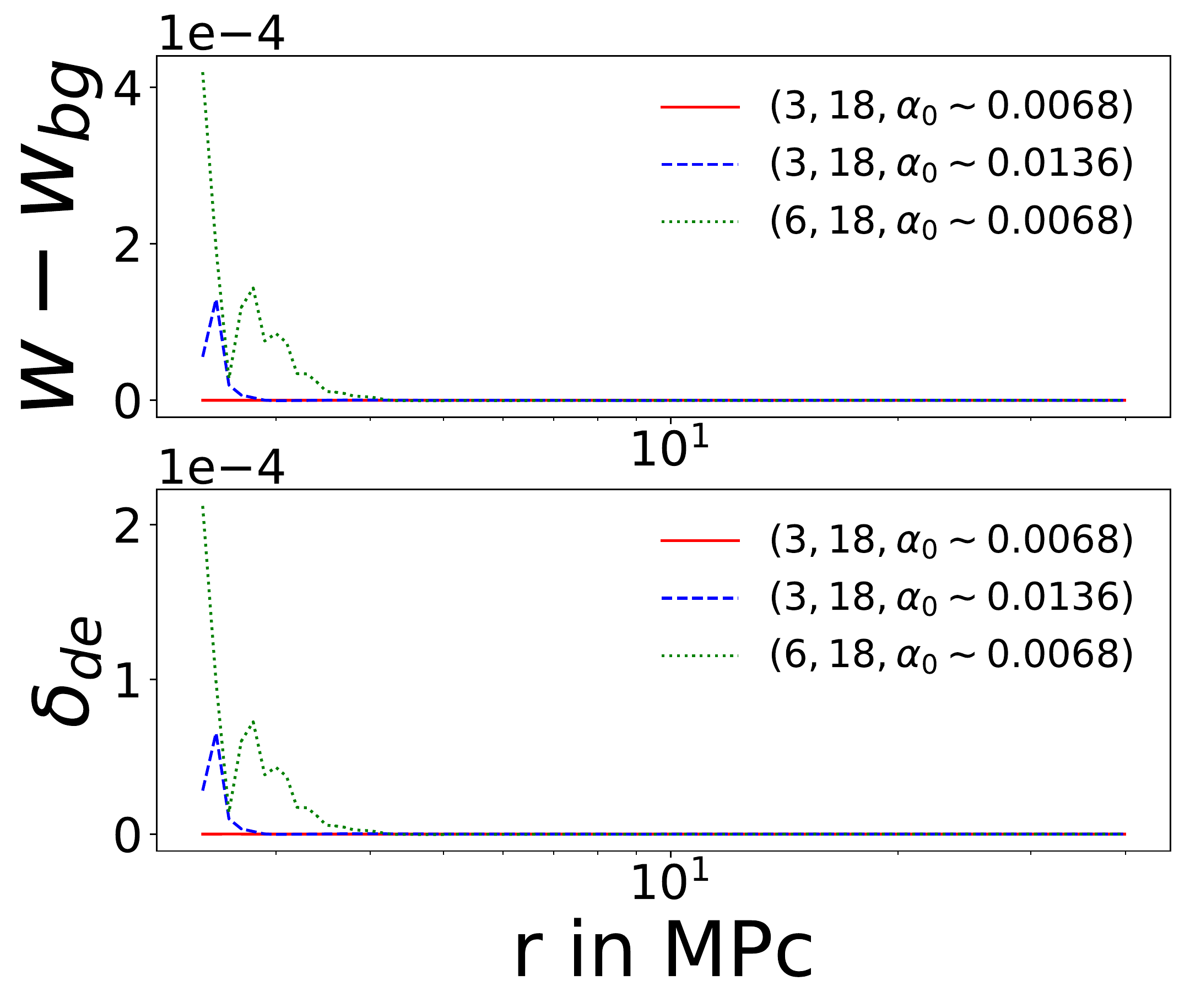}
    \includegraphics[width=0.45\textwidth]{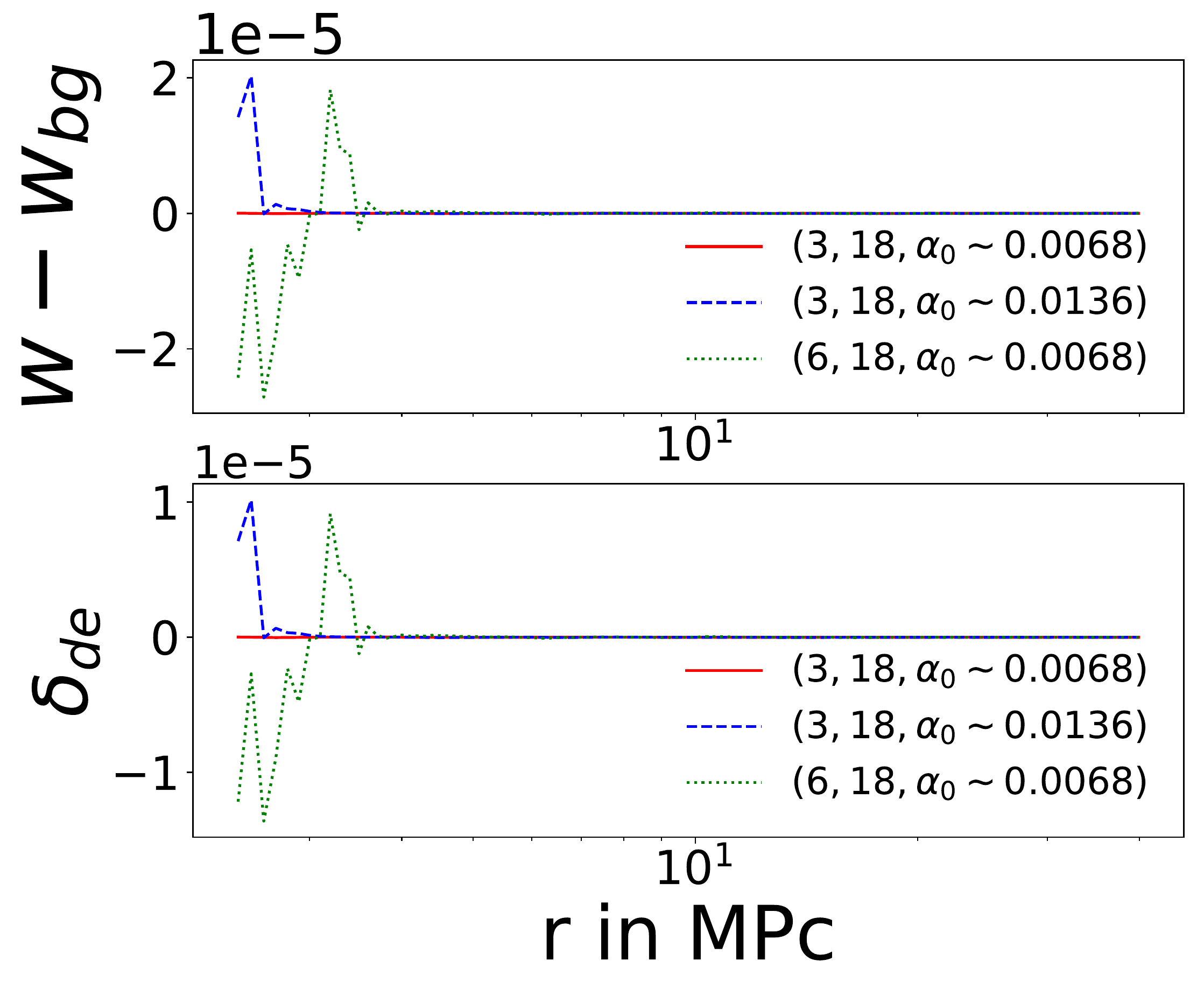}
    \caption{Effects of scale for small over dense case $(3,18)$. The
      fluctuations here are very weak and comparable to numerical
      noise. But they are suggestive of claims of made in
      article. Following figures clearly establish the claim that
      length scale of perturbations play a more important role.} 
    \label{fig:12}
\end{figure}

\begin{figure}
\renewcommand{\thefigure}{14(extra)}
    \centering
    \includegraphics[width=0.48\textwidth]{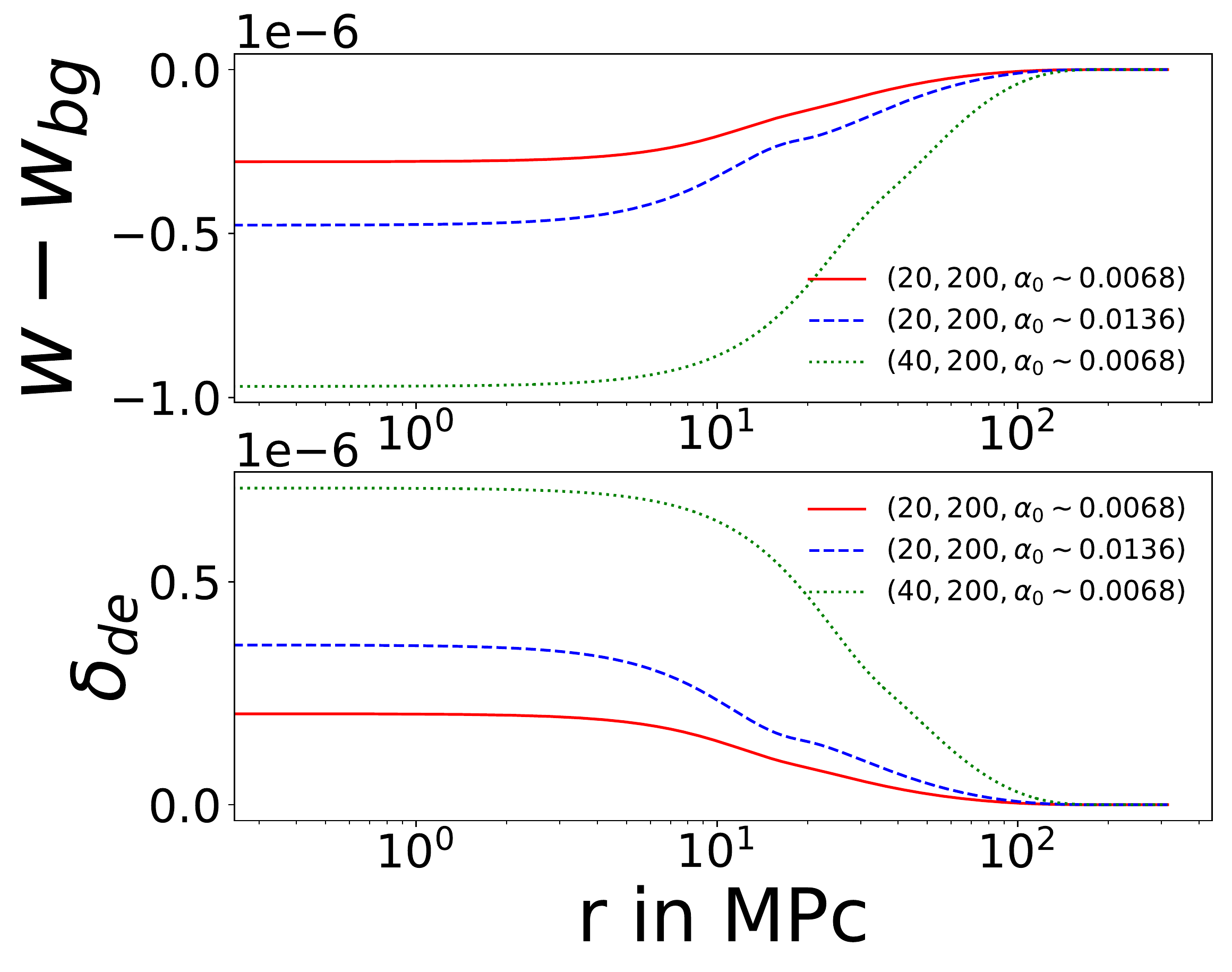}
    \includegraphics[width=0.45\textwidth]{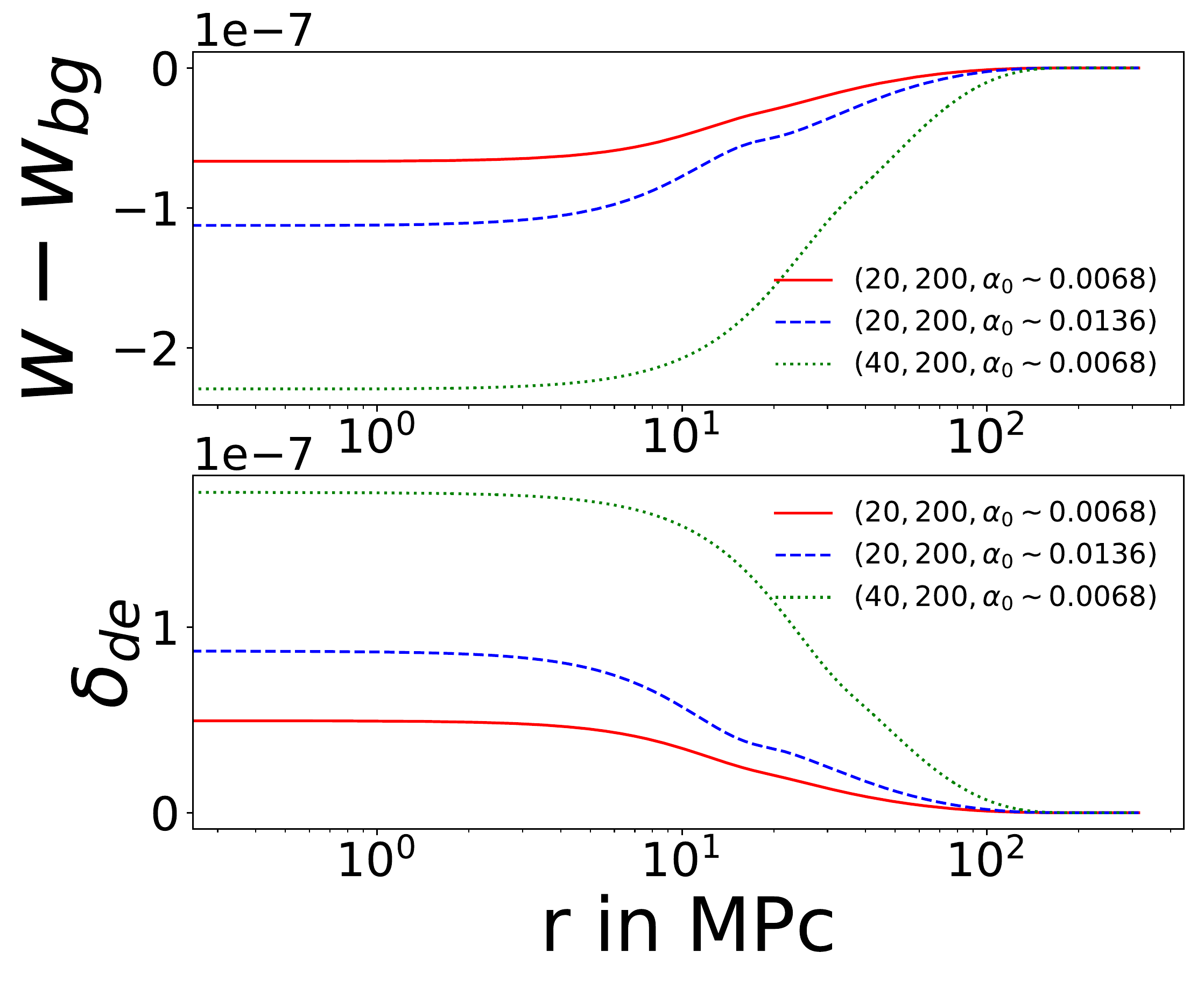}
    \caption{Effects of scale and amplitude variation for under dense
      regions.}  
    \label{fig:12u}
\end{figure}

We had demonstrated that the effect of scales are dominant when
compared with the effect of amplitude of perturbations.
This erratum in itself demonstrates this point. 
Variation for the over dense case presented in the original paper is
small but suggestive, see fig~(\ref{fig:12}).
The variation is much clearer for under dense perturbation as it is at
a larger scale, see fig~(\ref{fig:12u}).

\section{Effect of deviation in backgrounds}

\begin{figure}
    \centering
    \includegraphics[width=0.45\textwidth]{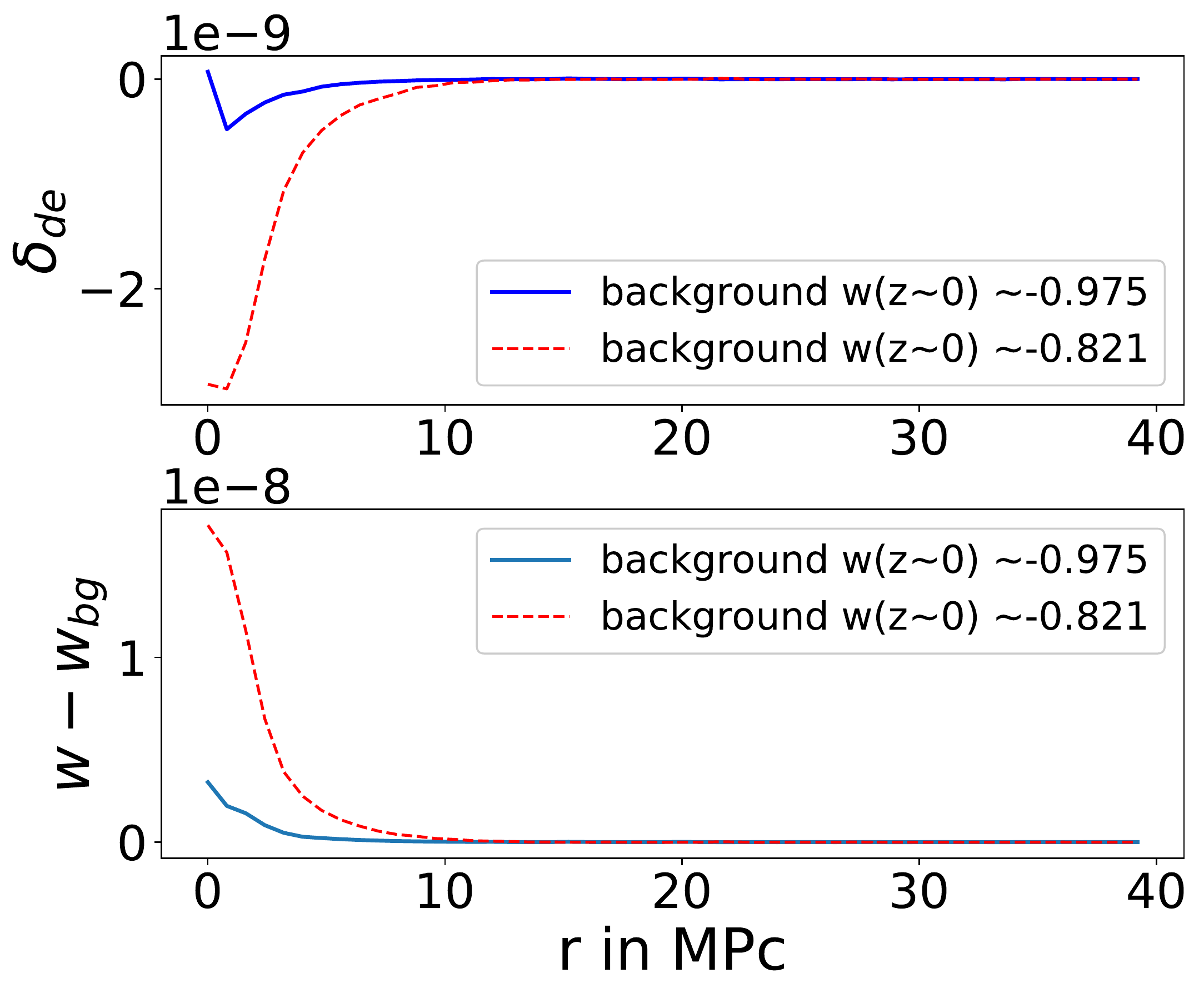}
    \includegraphics[width=0.45\textwidth]{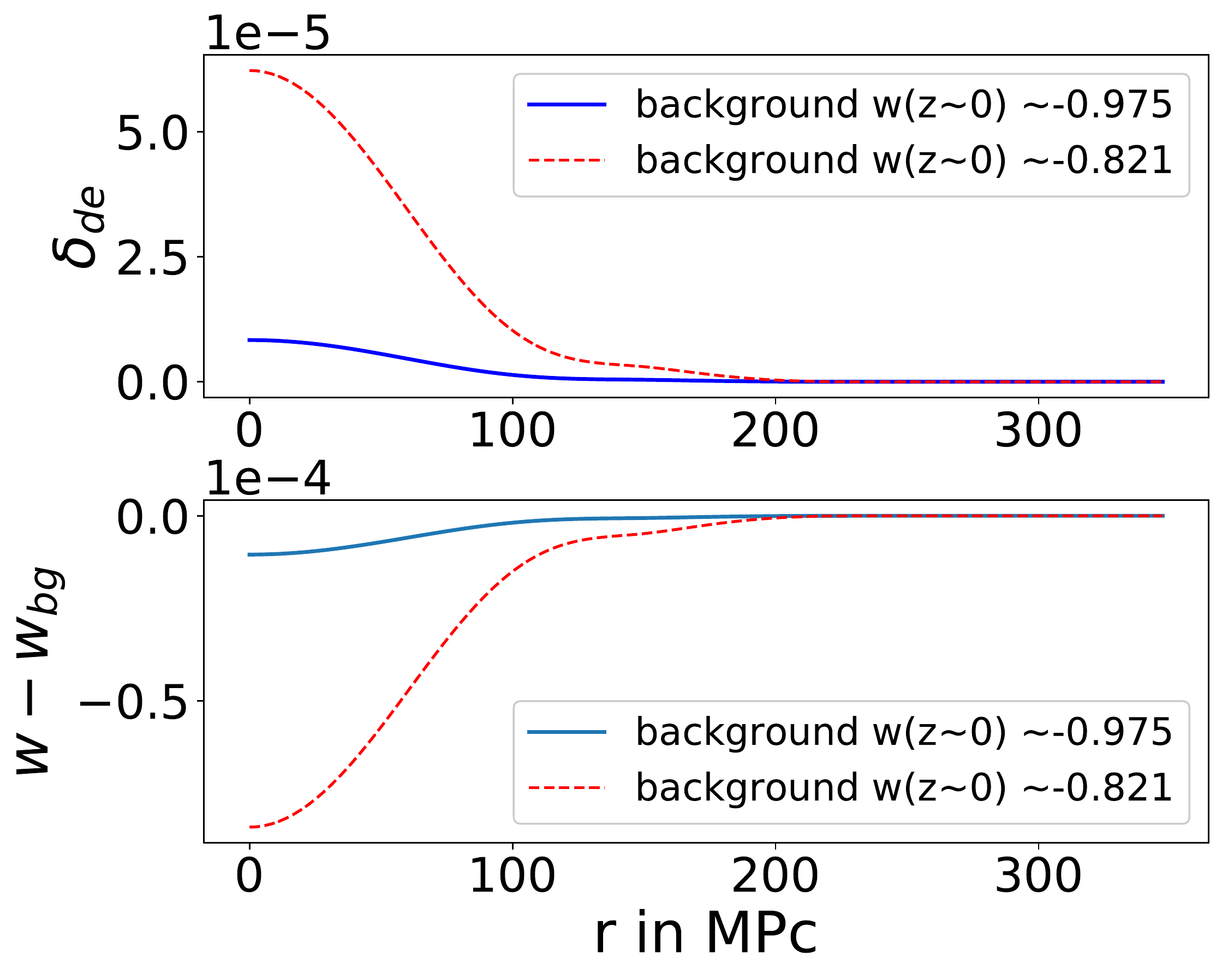}
    \caption{In this figure we study the impact of the equation of
      state parameter $w$ for the background on the growth of dark
      energy perturbations and the radial variation of the equation of
      state parameter.
      Here we show perturbation growth in two different background
      models for $V\propto \psi^2$.
      Curves are labeled by the present day values of $w$ for the
      background model.
      We see that the perturbations have a larger amplitude and $w$
      has a larger variation for a larger $1 + w_0$.
      The left panel here is for an over-density (OD1) and the right
      panel is for a void (UD5).
      We see that the effect is strongly pronounced for under density
      partly because of the larger scale of perturbation.
      The curves for over density are plotted for $z = 1.5$, before
      virialisation of the innermost shells. Curves for UD5 are
      plotted at $z = 0$.} 
    \label{fig:15ofa}
\end{figure}

We had argued that perturbations in dark energy grow by a significant
amount for models where $(1+w)$ is larger.
This continues to hold with the corrected scaling, as can  be seen in
fig~(\ref{fig:15ofa}).

\section{Virialization condition for the field}

In our work we considered three possibilities for dynamics of the
scalar field dynamics in the virialized region.
These are:
\begin{enumerate}
\item
  The scalar field can be evolved as a test field in the space-time
  determined by the frozen metric coefficients in the virialised
  region.
\item
  The scalar field can also be frozen in the virialised region, i.e.,
  we put $\dot{\psi} = 0 = \ddot{\psi}$ in this region.
\item
  We put $\ddot\psi=0$ and freeze the value of $\dot\psi(r)$ inside
  the virial region.
\end{enumerate}
We reported that ``the differences between 3 approaches decrease
rapidly beyond the turn around scale.''
We find that while working with corrected scales, the second and 3rd
approach show numerical instability.
We use the stable and consistent approach, the first one, which was
used in original article as well.

\bibliographystyle{plain}

\end{document}